\documentclass[10pt,conference]{IEEEtran}
\IEEEoverridecommandlockouts
\usepackage{cite}
\usepackage{amsmath,amssymb,amsfonts}
\usepackage{algorithmic}
\usepackage{graphicx}
\usepackage{textcomp}
\usepackage{xcolor}

\def\BibTeX{{\rm B\kern-.05em{\sc i\kern-.025em b}\kern-.08em
    T\kern-.1667em\lower.7ex\hbox{E}\kern-.125emX}}

\usepackage{url}
\usepackage{hyperref}
\hypersetup{hidelinks,
    colorlinks=true,
    allcolors=black,
    pdfstartview=Fit,
    breaklinks=true}
\usepackage{subfigure}
\usepackage{multirow}
\usepackage{makecell}
\usepackage{enumitem}
\newcommand{\ignore}[1]{}
\newtheorem{insight}{Insight}

\begin{document}

\title{Robin: A Novel Method to Produce Robust Interpreters for Deep Learning-Based Code Classifiers
\thanks{\IEEEauthorrefmark{2}National Engineering Research Center for Big Data Technology and System, Services Computing Technology and System Lab, Hubei Key Laboratory of Distributed System Security, Hubei Engineering Research Center on Big Data Security, Cluster and Grid Computing Lab}
\thanks{{\IEEEauthorrefmark{1}Corresponding author}}
}

\author{
\IEEEauthorblockN{
Zhen Li$^{1}$\IEEEauthorrefmark{2},
Ruqian Zhang$^{1}$\IEEEauthorrefmark{2},
Deqing Zou$^{1}$\IEEEauthorrefmark{2}\IEEEauthorrefmark{1},
Ning Wang$^{1}$\IEEEauthorrefmark{2},
Yating Li$^{1}$\IEEEauthorrefmark{2},
Shouhuai Xu$^{2}$,
Chen Chen$^{3}$,
and Hai Jin$^{4}$\IEEEauthorrefmark{2}
}
\IEEEauthorblockA{$^{1}$\textit{School of Cyber Science and Engineering, Huazhong University of Science and Technology, Wuhan, 430074, China}}
\IEEEauthorblockA{$^{2}$\textit{Department of Computer Science, University of Colorado Colorado Springs, USA}}
\IEEEauthorblockA{$^{3}$\textit{Center for Research in Computer Vision, University of Central Florida, USA}}
\IEEEauthorblockA{$^{4}$\textit{School of Computer Science and Technology, Huazhong University of Science and Technology, Wuhan, 430074, China}}
\IEEEauthorblockA{\{zh\_li, ruqianzhang, deqingzou, wangn, leeyating\}@hust.edu.cn \\
sxu@uccs.edu, chen.chen@crcv.ucf.edu, hjin@hust.edu.cn}
}

\maketitle

\begin{abstract}
Deep learning has been widely used in source code classification tasks, such as code classification according to their functionalities, code authorship attribution, and vulnerability detection. Unfortunately, the black-box nature of deep learning makes it hard to interpret and understand why a classifier (i.e., classification model) makes a particular prediction on a given example. This lack of interpretability (or explainability) might have hindered their adoption by practitioners because it is not clear when they should or should not trust a classifier's prediction.
The lack of interpretability has motivated a number of studies in recent years. However, existing methods are neither robust nor able to cope with out-of-distribution examples.
In this paper, we propose a novel method to produce \underline{Rob}ust \underline{in}terpreters  
for a given deep learning-based code classifier; the method is dubbed Robin.  
The key idea behind Robin is a novel hybrid structure combining an interpreter and two approximators, while leveraging
the ideas of adversarial training and data augmentation.
Experimental results show that on average the interpreter produced by Robin achieves a 6.11\% higher fidelity (evaluated on the classifier),
67.22\% higher fidelity (evaluated on the approximator), and 
15.87x higher robustness than that of the three existing interpreters we evaluated. 
Moreover, the interpreter is 47.31\% less affected by out-of-distribution examples than that of LEMNA.
\end{abstract}

\begin{IEEEkeywords}
Explainable AI, deep learning, code classification, robustness
\end{IEEEkeywords}

\section{Introduction}
In the past few years there has been an emerging field focusing on leveraging deep learning or neural networks to study various kinds of source code classification problems, such as classifying code based on their functionalities \cite{zhang2019novel,mou2016convolutional}, code authorship attribution \cite{caliskan2015anonymizing, alsulami2017source,yang2017authorship,bogomolov2021authorship,abuhamad2018large}, and vulnerability detection \cite {lin2017poster,li2018vuldeepecker, li2021sysevr}.  
While the accuracy of deep neural networks in this field may be satisfactory, the lack of interpretability, or explainability, remains a significant challenge. Deep neural networks are often considered {\em black-boxes} which means they cannot provide explanations for why a particular prediction is made. 
The lack of interpretability poses as a big hurdle to the adoption of these models in the real world (particularly in high-security scenarios), because practitioners do not know when they should trust the predictions made by these models and when they should not.

The importance of addressing the aforementioned lack of interpretability is well recognized by the research community\cite{vaswani2017attention, choi2016retain, ribeiro2016should}, as evidenced by very recent studies. Existing studies on addressing the interpretability of source code classifiers (i.e., classification models)
can be classified into two approaches: {\em ante-hoc} vs. {\em post-hoc}. 
The ante-hoc approach aims to provide {\em built-in} interpretability by leveraging 
the attention weight matrix associated with a neural network in question \cite{bui2019autofocus, zou2022mvulpreter}, which in principle can be applied to explain the prediction on any example. The post-hoc approach 
aims to interpret the decision-making basis of a trained model. 
In the context of source code classification, this approach mainly focuses on local interpretation, which aims to explain predictions for individual examples
by leveraging: (i) perturbation-based feature saliency \cite{zou2021interpreting, cito2022counterfactual}, which computes the importance scores of features by perturbing features in code examples and then observing changes in prediction scores; or (ii) program reduction \cite{suneja2021probing, rabin2021understanding}, which uses the delta debugging technique \cite{zeller2002simplifying} to reduce a program to a minimal set of statements while preserving the classifier's prediction. 

The ante-hoc approach must be incorporated into the classifier training phase, meaning that it cannot help existing or given classifiers, for which we can only design interpreters to provide interpretability in a retrospective manner. 
In this paper we focus on how to retrospectively equip given code classifiers with interpretability, which is the focus of the post-hoc approach.
However, existing post-hoc methods suffer from the following two problems.
(i) The {\em first} problem is incurred by the possible out-of-distribution of a perturbed example in the perturbation-based feature saliency method. This is inevitable because the method uses perturbations to assess feature importance, by identifying the feature(s) whose absence causes a significant decrease in prediction accuracy. When a legitimate example is perturbed into an out-of-distribution input, it is unknown whether the drop in accuracy is caused by the absence of certain feature(s) or because of the out-of-distribution of the perturbed example \cite{hooker2019benchmark,brocki2022evaluation}. 
(ii) The {\em second} problem is the lack of robustness, which is inherent to the local interpretation approach and thus common to both the perturbation-based feature saliency method and the program reduction method. 
This is because the local interpretation approach optimizes the interpretation of each example independent of others,
meaning that overfitting the noise associated with individual examples is very likely \cite{bajaj2021robust}. As a consequence, an interpretation would change significantly even by incurring a slight modification to an example, and this kind of sensitivity could be exploited by attackers to ruin the interpretability 
\cite{zhang2020interpretable}.
The weaknesses of the existing approaches motivate us to investigate better methods  
to interpret the predictions of deep learning-based code classifiers. 

\smallskip
\noindent {\bf Our Contributions.}
This paper introduces Robin, a novel method for producing high-fidelity and \underline{Rob}ust \underline{in}terpreters in the post-hoc approach with local interpretation. Specifically, this paper makes three contributions.

First, we address the aforementioned out-of-distribution problem by introducing a hybrid interpreter-approximator structure. More specifically, we design (i) an interpreter to identify the features that are important to make accurate predictions, and (ii) two approximators such that one is used to make predictions based on these important features and the other is used to make predictions based on the other features (than the important ones). 
These approximators are reminiscent of fine-tuning a classifier with perturbed training examples while removing some features.
As a result, a perturbed test example is no longer an out-of-distribution example to the approximators, meaning that the reduced accuracy of the classifier can be attributed to the removal of features (rather than the out-of-distribution examples). To assess the importance of the features extracted by the interpreter, we use the approximators (rather than the given classifier) to mitigate the side-effect that may be caused by out-of-distribution examples.

Second, we address the lack of interpretation robustness by leveraging the ideas of {\em adversarial training} and {\em mixup} to augment the training set.
More specifically, we generate a set of perturbed examples for a training example (dubbed the {\em original} example) as follows. 
(i) Corresponding to adversarial training but different from traditional adversarial training in other contexts, the ground-truth labels (i.e., what the $k$ important features are) cannot be obtained, making it difficult to add perturbed examples to the training set for adversarial training. We overcome this by measuring the similarity between the interpretation of the prediction on the original example and the interpretation of the prediction on the perturbed example, which is obtained in the example space rather than feature space (i.e., the perturbed example is still a legitimate program with the same functionality as the original example). This similarity allows us to compute a loss in interpretability and leverage this loss to train  
the interpreter. 
(ii) Corresponding to mixup, 
we generate a set of {\em virtual} examples by linearly interpolating  
the original examples and their perturbed versions in the feature space; these examples are {\em virtual} because they are obtained in the feature space (rather than example space) and thus may not correspond to any legitimate code example (e.g., a virtual example may not correspond to a legitimate program). Different from traditional data augmentation, we train the interpreter and two approximators jointly rather than solely training the interpreter on virtual examples due to the lack of ground truth of the virtual examples (i.e., what the $k$ important features are).

Third, we empirically evaluate Robin's effectiveness and compare it with the known post-hoc methods in terms of {\em fidelity}, {\em robustness}, and {\em effectiveness}.
Experimental results show that on average the interpreter produced by Robin achieves a 6.11\% higher fidelity (evaluated on the classifier), 67.22\% higher fidelity (evaluated on the approximator), and 15.87x higher robustness than that of the three existing interpreters we evaluated. 
Moreover, the interpreter is 47.31\% less affected by out-of-distribution examples than that of LEMNA\cite{guo2018lemna}. 
We have made the source code of Robin publicly available at \textcolor{magenta}{{https://github.com/CGCL-codes/Robin}}.

\smallskip
\noindent
{\bf Paper Organization.} Section \ref{sec:motivation} presents a motivating instance. Section \ref{sec:design} describes the design of Robin. Section \ref{sec:experiments} presents our experiments and results. Section \ref{sec:limitations} discusses the limitations of the present study. Section \ref{sec:related_work} reviews related prior studies. Section \ref{sec:conclusion} concludes the paper.

\section{A Motivating Instance}
\label{sec:motivation}
To illustrate the aforementioned problem inherent to the local interpretation approach, we consider a specific instance of code classification in the context of code functionality classification via TBCNN \cite{mou2016convolutional}. Although TBCNN offers code functionality classification capabilities, it does not offer any interpretability on its predictions. To make its predictions interpretable, an interpreter is required.
We adapt the method proposed in \cite{zou2021interpreting}, which was originally used to interpret software vulnerability detectors, to code functionality classification because there are currently no existing interpreters for this purpose (to the best of our knowledge). This adaptation is feasible because the method involves deleting features in the feature space and observing the impact on predictions, which is equally applicable to code functionality classification.

\begin{figure}[!tb]
  \centering
  \vspace{-0.2cm}
  \subfigure[Original example]{
    \begin{minipage}{0.46\linewidth} 
    \centering
\includegraphics[height=1.5\textwidth]{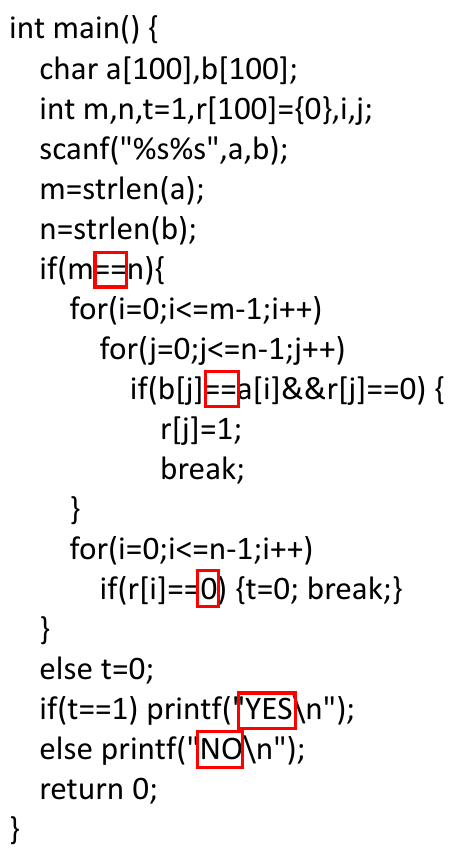} 
    \end{minipage}
	}
  \subfigure[Perturbed example]{
    \begin{minipage}{0.46\linewidth}
    \centering
      \includegraphics[height=1.5\textwidth]{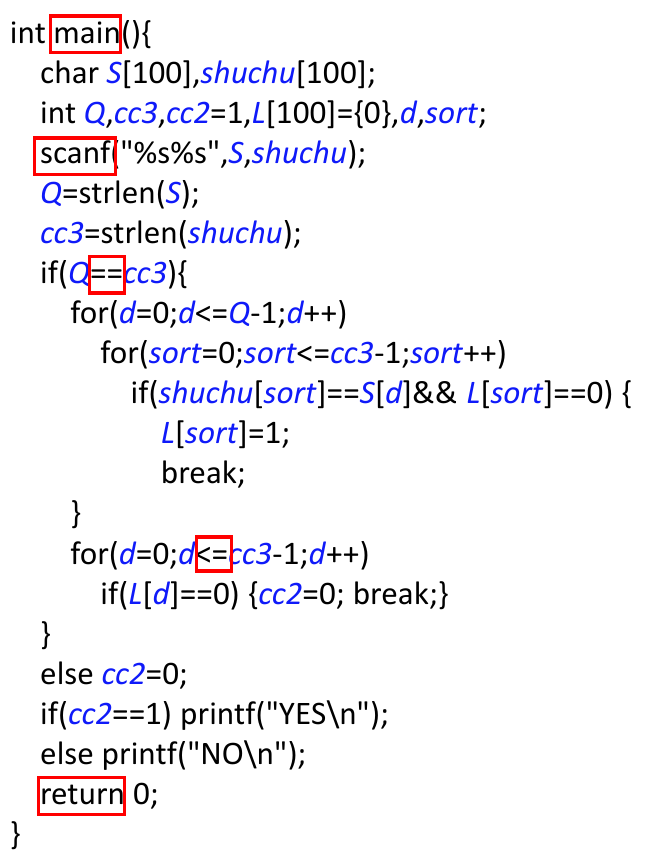}
    \end{minipage}
	}
 \vspace{-0.2cm}
  \caption{An original example and its perturbed example (modified code is highlighted in blue color and italics), where red boxes highlight the 5 most important features.}
  \label{fig:code}
  \vspace{-0.6cm}
\end{figure}

\begin{figure*}[!tb]
\vspace{-0.2cm}
  \centering
  \includegraphics[width=.8\linewidth]{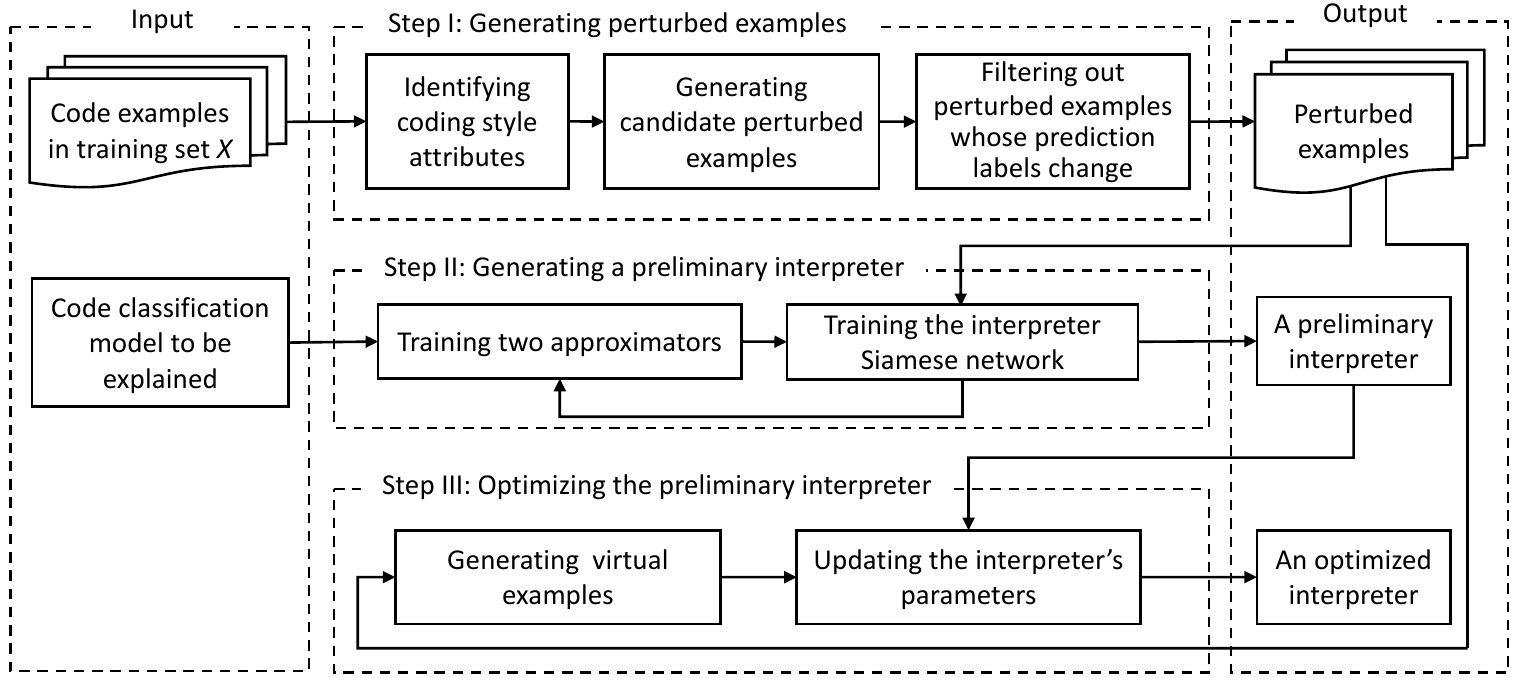}
  \vspace{-0.3cm}
  \caption{Overview of the training process of Robin which produces an optimized interpreter in three steps: generating perturbed examples, generating a preliminary interpreter, and optimizing the preliminary interpreter.}
  \label{fig:framework}
  \vspace{-0.2cm}
\end{figure*}

The original code example in Figure \ref{fig:code} (a) is used to compare two strings for equality. We create a perturbed example by changing the variable names, as illustrated in Figure \ref{fig:code} (b). Despite the change in variable names, the perturbed example maintains the same functionality and semantics as the original example. Additionally, both the original and perturbed examples are classified into the same category by the classifier.
Upon applying an interpreter, adapted from \cite{zou2021interpreting}, to TBCNN, the five most important features of the original and perturbed examples are identified, and highlighted in Fig.~\ref{fig:code}(a) and \ref{fig:code}(b) respectively. Notably, only one important feature is common between the two examples, revealing that the interpreter lacks robustness. 
This lack of robustness of the interpreter may cause users to question the reliability of the classifier's predictions 
due to the erroneous interpretation.

\section{Design of Robin}
\label{sec:design}

\noindent{\bf Notations}. A program code example, denoted by $x_i$, 
can be represented as a $n$-dimensional feature vector $x_i=(x_{i,1},x_{i,2}, \ldots, x_{i,n})$,
where $x_{i,j}$ ($1\le{j}\le{n}$) is the $j$th feature 
of $x_i$. 
A code classifier (i.e., classification model) $M$ is learned from a training set, denoted by $X$,
where each example $x_i \in X$ 
is associated with a label $y_i$. 
Denote by $M(x_i)$ the prediction of 
classifier $M$ on a example $x_i$.

Our goal is to propose a novel method to produce an interpreter, denoted by $E$, for {\em any} given code classifier $M$ and test set $U$ such that for test example ${u}_i\in U$,
$E$ identifies $k$ important features to explain why $M$ makes a particular prediction on ${u}_i$, where $k\ll{n}$.
It is intuitive that the $k$ important features of example $u_i$ should be largely, if not exactly, the same as the $k$ important features of $u'_i$ which is perturbed from $u_i$. 
Denote by $E(u_i)=(u_{i,\alpha_1},... ,u_{i,\alpha_k})$ the $k$ important features identified by $E$, where $\{\alpha_1,...,\alpha_k\}\subset\{1, \ldots,n\}$.

\begin{figure*}[!tb]
\vspace{-0.2cm}
  \centering
  \includegraphics[width=\linewidth]{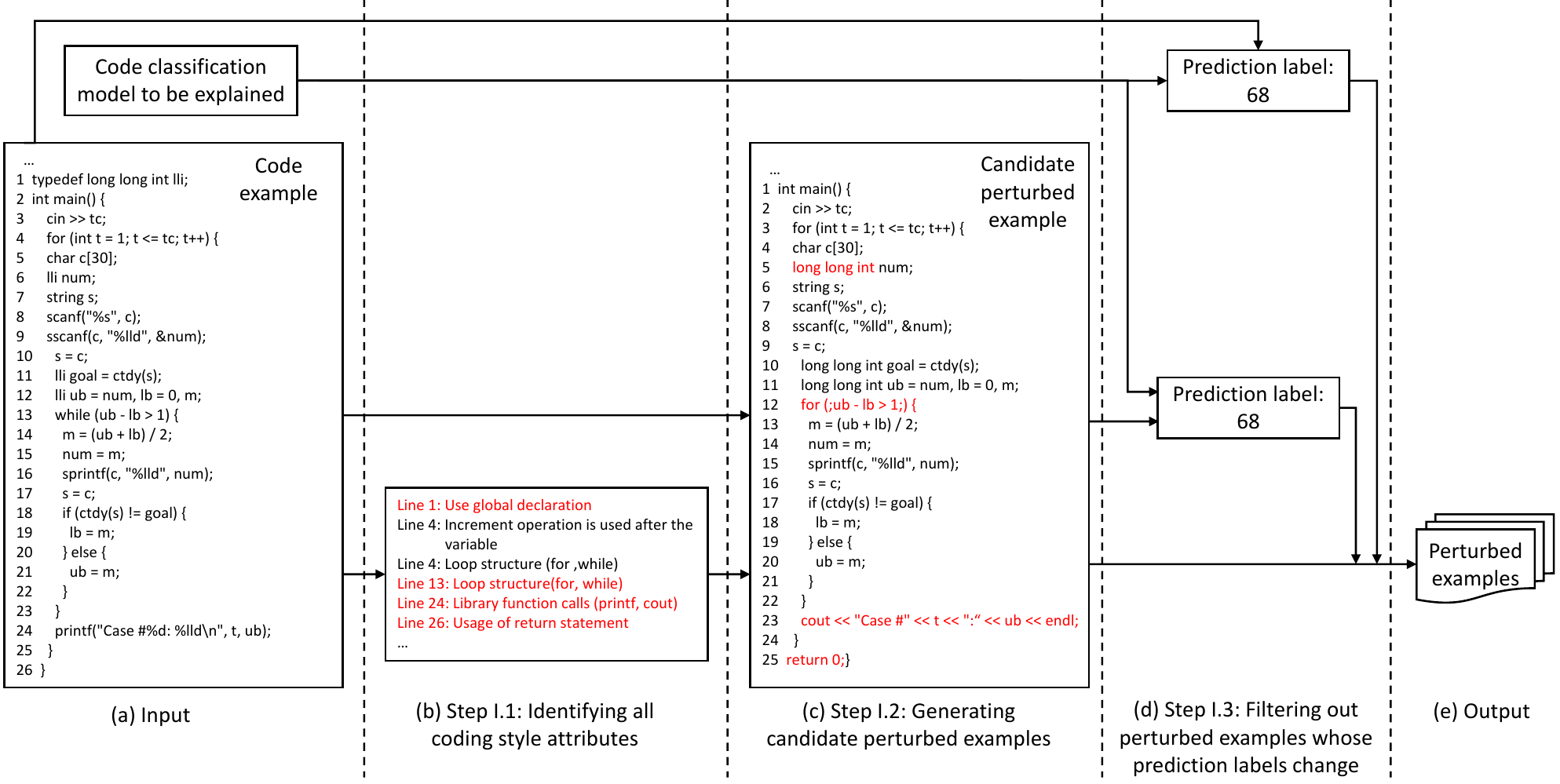}
  \vspace{-0.6cm}
  \caption{A code example showing generation of perturbed examples (selected coding style attributes and modified code are highlighted in red).}
  \label{fig:example}
  \vspace{-0.5cm}
\end{figure*}

\subsection{Basic Idea and Design Overview}
\noindent
{\bf Basic Idea.} 
In terms of the out-of-distribution problem associated with existing interpretation methods, we observe that the absence of perturbed examples in the training set makes a classifier's prediction accuracy with respect to the perturbed examples affected by the out-of-distribution examples. Our idea to mitigate this problem is to fine-tune a classifier for perturbed examples by using a hybrid interpreter-approximator structure \cite{chen2018learning} such that (i) one interpreter is for identifying the important features for making accurate prediction, (ii) one approximator is for using the important features (identified by the interpreter) to making predictions, and (iii) another approximator is for using the other features (than the important features) for making predictions. To improve the interpreter's fidelity, the two approximators are trained simultaneously such that the important features contain the most useful information for making predictions while the other features contain the least useful information for making predictions.

To make the interpreter robust, we leverage two ideas. The {\em first idea} is to
use adversarial training \cite{lakkaraju2020robust, la2021guaranteed} where an original example and its perturbed example will have the same prediction. In sharp contrast to traditional adversarial training in other contexts where ground-truth can be obtained, it is difficult to obtain the ground-truth labels in this setting because we do not know which features are indeed the most important ones even for the training examples.
That is, we cannot simply use traditional adversarial training method to add perturbed examples to training set because the ``labels'' (i.e., the $k$ important features) of original examples and perturbed examples cannot be obtained. 
We overcome this by
(i) generating a set of perturbed examples via code transformation such that the prediction on the perturbed example remains the same, and (ii) adding a constraint term to the loss function to make the interpretations of the original example and the perturbed example as similar to each other as possible.

The {\em second idea} is to 
leverage mixup \cite{zhang2017mixup} to augment the training set. 
 In sharp contrast to traditional data augmentation, we cannot train the interpreter from the augmented dataset for the lack of ground-truth (i.e., the important features of an original example and its perturbed examples can not be obtained). 
We overcome this issue by (i) using code transformation 
to generate a perturbed example such that its prediction remains the same as that of the original example,  
(ii) mixing the original examples with the perturbed examples to generate virtual examples, and (iii) optimizing the preliminary interpreter by training the interpreter and two approximators jointly on virtual examples. Note that the difference between the aforementioned adversarial examples and virtual examples is that the former are obtained by perturbation in the example space but the latter is obtained in the feature space.

\smallskip

\noindent{\bf Design Overview}.
Fig.~\ref{fig:framework} highlights the training process of Robin, which produces an optimized interpreter in three steps.
\begin{itemize}
[leftmargin=.32cm,noitemsep,topsep=2pt]
\item {\bf Step I: Generating perturbed examples.} This step generates perturbed examples from a training example by conducting semantics-preserving code transformations such that the perturbed example has the same prediction as that of the original example. 
\item {\bf Step II: Generating a preliminary interpreter.} Given a classifier for which we want to equip with interpretability, this step leverages the perturbed examples generated in Step I to train two approximators and an interpreter Siamese network in an iterative fashion. The interpreter Siamese network identifies the important features of original examples and that of their perturbed examples, and then computes the difference between these two sets.
\item {\bf Step III: Optimizing the preliminary interpreter.} This step optimizes the preliminary interpreter generated in Step II by using mixup \cite{zhang2017mixup} to augment the training set and update the preliminary interpreter's parameters. 
The optimized interpreter identifies important features of a test example.
\end{itemize}

\begin{figure*}[!tb]
\vspace{-0.2cm}
  \centering
  \includegraphics[width=\linewidth]{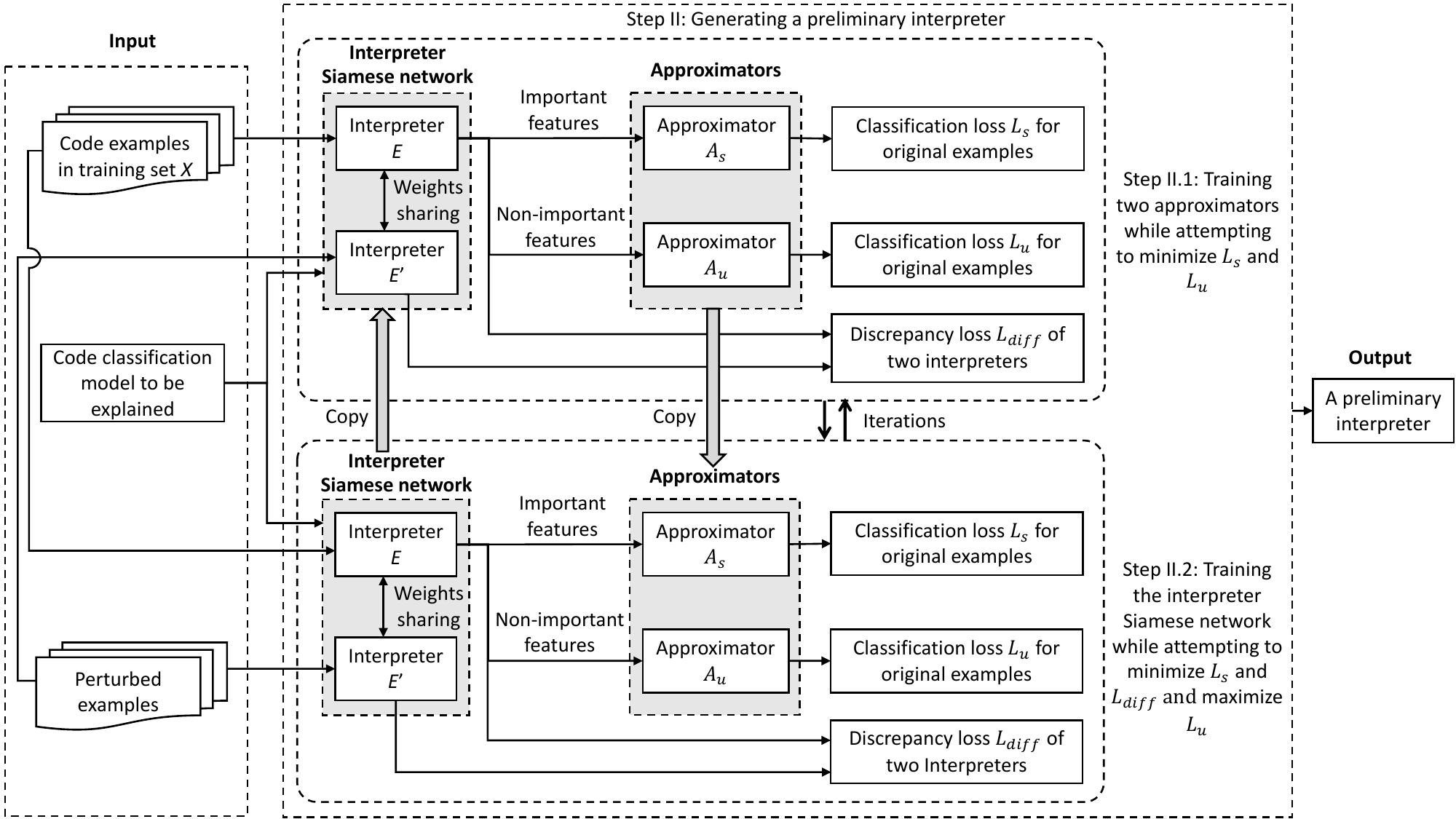}
  \vspace{-0.7cm}
  \caption{Overview of Step II (generating a preliminary interpreter), involving training two approximators and training the interpreter Siamese network iteratively.}
  \label{fig:explainer}
  \vspace{-0.5cm}
\end{figure*}

\subsection{Generating Perturbed Examples 
} 
This step has three substeps. First, for each training example $x_i\in X$, we identify its
coding style attributes related to the example's layout, lexis, syntax, and semantics (e.g., as defined in \cite{li2022ropgen}).
Let $t_i$ denote the number of coding style attributes for $x_i\in X$.
Fig.~\ref{fig:example}(a) shows a training example 
where the first line  
uses a global declaration but can be transformed such that no global declaration is used; Fig.~\ref{fig:example}(b) describes its coding style attributes. 
Second, we randomly select $\theta_i$ ($\theta_i < t_i$) coding style attributes, repeat this process for $m$ times, and transform 
the value of each of these
coding style attributes to any one of the other semantically-equivalent coding style attributes. 
Consequently, we obtain $m$ candidate perturbed examples $x_{i,+1},x_{i,+2}... .x_{i,+m}$, where $x_{i,+j}$ ($1 \leq j \leq m$) denotes the $j$th perturbed example generated by the semantic-equivalent code transformation of code example $x_i$. 
The labels of the $m$ perturbed examples preserve the original example $x_i$'s label $y_i$ owing to the semantic-equivalent code transformations.
As an instance, 
Fig.~\ref{fig:example}(c) shows a candidate perturbed example generated by transforming randomly selected coding style attributes, which 
are highlighted in red in Fig.~\ref{fig:example}(b).
Third, we filter out perturbed examples whose prediction labels are different from the prediction labels of the corresponding original examples in $X$. The reason is that if the prediction labels of the perturbed examples change, the robustness of the interpreter cannot be judged by the difference of the interpretation between the perturbed examples and the original examples.  
As an instance
in Fig.~\ref{fig:example}(d), the prediction label of the code example in Fig.~\ref{fig:example}(a) and the prediction label of the candidate perturbed example are the same, so the candidate perturbed example is a perturbed example that can be used for robustness enhancement.
Finally, we obtain the set of perturbed examples  
for robustness enhancement of the interpreter.

\subsection{Generating a Preliminary Interpreter 
}

An ideal interpreter is simultaneously achieving  
high fidelity and high robustness. 
 (i) {\em High fidelity} indicates that 
the important features identified by interpreter $E$ contain as much information as possible that is most useful for code classification, and the remaining non-important features contain as little information as possible that is useful for code classification. 
 (ii) {\em High robustness} indicates that 
the important features identified by interpreter $E$ to explain why $M$ predicts $x_i$ as the label $y_i$ 
should not change dramatically for small perturbed examples which are predicted as label $y_i$.
Robin achieves this  
by first generating a preliminary interpreter in Step II and then optimizing the preliminary interpreter further in Step III.

The purpose of Step II is to generate a preliminary interpreter by training two approximators and the interpreter Siamese network iteratively. The basic idea is as follows:
(i) To achieve high fidelity, we introduce two approximators that have the same neural network structure 
for code classification, using
the identified important features and non-important features as input respectively.  
Since important features contain the most useful information for code classification, the accuracy of the approximator using only  important features as input
should be as high as possible. On the other hand, non-important features contain less information important for code classification, so the accuracy of the approximator using only non-important features as input should be as low as possible. 
(ii) To achieve high robustness, we introduce the interpreter Siamese network with two interpreters that have the same neural network structure and share weights, using the original code examples and perturbed examples as input respectively.   
For each original example and its corresponding perturbed examples, the Siamese network calculates the similarity distance between the important features of the original example and the important features of the perturbed examples identified by the two interpreters, and adds the similarity distance to the loss value to improve the interpreters' robustness during training.

Fig.~\ref{fig:explainer} shows the structure of the neural network involving an interpreter Siamese network and two approximators. 
The interpreter Siamese network involves two interpreters which have the same neural network structure and share weights. Their neural network structure depends on the structure of the code classifier to be explained.
We divide the code classifier to be explained into two parts. One part is to extract the features from the input code examples through neural network to obtain the vector representation of the code examples, which is equivalent to encoder, and this part usually uses Batch Normalization, Embedding, LSTM, Convolutional layer, etc. The other part maps the vector representation to the output vector. When generating the structure of the interpreter, the first part of the code classifier is kept and the latter part is modified to a fully connected layer and a softmax layer, which maps the learned representation of code examples to the output space, and the output is of the same length as the number of features, indicating whether each feature is labeled as important or not.
These two interpreters are used to identify the important features of the code examples in training set $X$ and the perturbed examples generated in Step I, respectively. 

The two approximators have the same neural network structure  
and are used to predict labels using important features and non-important features, respectively. 
They have the identical neural network architecture as the code classifier to be interpreted. However, instead of the code example as input, the interpreter provides the approximator with the important or non-important features identified. As a result, the approximators can be seen as fine-tuned versions of the code classifier, trained on the datasets of important and non-important features.

Fig.~\ref{fig:explainer} also shows the training process to generate a preliminary interpreter, involving the following two substeps.

\noindent
{\bf Step II.1: Training two approximators while attempting to minimize $L_s$ and $L_u$.}
When training the approximator, only the model parameters of the approximator are updated. The training goal is to minimize the loss of both approximators $A_s$ and $A_u$, which is the sum of cross-entropies loss of $A_s$ and $A_u$: 

\vspace{-0.4cm}
\begin{equation}
\min_{A_s,A_u}{(L_s+L_u)},
\label{eq2}
\end{equation}
where $L_s$ is the cross-entropy loss of approximator $A_s$ and $L_u$ is the cross-entropy loss of approximator $A_u$.
The loss of the approximator indicates the consistency between the prediction labels and the labels.

\noindent
{\bf Step II.2: Training the interpreter Siamese network while attempting to minimize $L_s$ and $L_{diff}$, and maximize $L_u$.}
When training the interpreter Siamese network, only the model parameters of the interpreter are updated. The training goal is to minimize the loss of $A_s$ and the discrepancy of the outputs between two interpreters $E$ and $E'$, and maximize the loss of $A_u$: 
\begin{equation}
\min_{E}{(L_s-L_u+L_{diff})},
\label{eq3}
\end{equation}
where $L_{diff}$ is the discrepancy of the outputs between two interpreters $E$ and $E'$.
The interpreter is trained so that (i)
the loss of prediction using important features is minimized, (ii) 
the loss of prediction using non-important features is maximized, and (iii) the discrepancy of the outputs between two interpreters is minimized to improve the robustness of the interpreter.
The difference value $L_{diff}$ in the interpreter Siamese network represents the distance between the important features identified by the interpreter for the original examples and those for the perturbed examples. We use Jaccard distance \cite{levandowsky1971distance}
to measure the distance as follow:
\begin{equation}
L_{diff}=1-\sum_{i,j}\frac{\left|{E(x_i)\cap{E(x_{i,+j})}}\right|}{N\cdot m\cdot \left|{E(x_i)\cup{E(x_{i,+j})}}\right|}
\label{eq1}
\end{equation}
where $N$ is the number of original code examples in the training set $X$, and $m$ is the number of perturbed examples corresponding to each original example.
The more robust the interpreter is, the higher the similarity between the important features for the original examples and for the perturbed examples, the smaller the Jaccard distance, and the smaller the corresponding difference value $L_{diff}$. 

During the training process, Step II.1 and Step II.2 are iterated 
until both the interpreters and the approximators converge.

\subsection{Optimizing the Preliminary Interpreter
}
The purpose of this step is to optimize the preliminary interpreter generated in Step II in both fidelity and robustness. The basic idea is to use mixup \cite{zhang2017mixup} for data augmentation to optimize the interpreter.
There are two substeps. First, we generate virtual examples. 
For each code example $x_i$ in training set $X$, $x_{i^{'},+j}$ is a randomly selected perturbed example of $x_{i'}$, where $x_{i'}$ is randomly selected from $X$, and may or may not be identical to $x_i$. 
A virtual example is generated by mixing code examples and their corresponding labels. Specifically, the virtual example $x_{i,mix}$ is generated by linear interpolation between the original example $x_i$ and the perturbed example $x_{i^{'},+j}$, and the label $y_{i,mix}$ of $x_{i,mix}$ is also generated by linear interpolation between the label $y_i$ of original example $x_i$ and the label $y_{i^{'},+j}$ of perturbed example $x_{i^{'},+j}$, shown as follows:
\begin{equation}
\begin{aligned}
& x_{i,mix}=\lambda_i x_i+(1-\lambda_i)x_{i^{'},+j} \\
& y_{i,mix}=\lambda_i y_i+(1-\lambda_i)y_{i^{'},+j}
\label{eq4}
\end{aligned}
\end{equation}
where the interpolation coefficients $\lambda_i$
is sampled from the $\beta$ distribution.
Second, we update the interpreter's parameters based on the generated virtual examples. 
Since the output of the interpreter is the important features in code examples rather than the classification labels, it is impossible to train the interpreter individually for enhancement. Therefore, we use approximators for joint optimization with the interpreter $E$. 
In this case, the input of the overall model are code examples and the output are the labels of code examples, which can be directly trained and optimized using the generated virtual examples. 
In the optimization process, 
the interpreter's parameters are updated while preserving
the approximators' parameters unchanged.

\section{Experiments and Results}
\label{sec:experiments}

\subsection{Evaluation Metrics and Research Questions}

\noindent {\bf Evaluation Metrics.}
We evaluate interpreters via their {\em fidelity}, {\em robustness} against perturbations, and {\em effectiveness} in coping with 
out-of-distribution examples.

For quantifying fidelity, we adopt the metrics defined in  
\cite{chen2018learning,liang2020adversarial}. 
Consider a code classifier $M$ trained from a training set $X$, an interpreter $E$, and a test set $U$. Denote by $E(u_i)$ the set of important features identified by interpreter $E$ for test example $u_i\in U$. We train an approximator $A_s$ in the same fashion as how $M$ 
is trained except that we only consider the important features, 
namely $\cup_{u_i\in U} E(u_i)$. 
Let $M(u_i)$ and $A_s(u_i)$ respectively denote the prediction of classifier $M$ and approximator $A_s$ on example $u_i$.
Then, interpreter $E$'s fidelity is defined as a pair (FS-M$\in[0,1]$, FS-A$\in[0,1]$), where FS-M$=\frac{|\{u_i\in U: M(u_i)=M(E(u_i))\}|}{|U|}$ is the fraction of test examples that have the same predictions by $M$ using all features and by $M$ only using the important features,
and FS-A$=\frac{|\{u_i\in U: M(u_i)=A_s(E(u_i))\}|}{|U|}$ is the fraction of test examples that have the same predictions by $M$ using all features and by $A_s$ only using the important features \cite{liang2020adversarial}. 
Note that a larger (FS-M, FS-A)  
indicates a higher fidelity, meaning that the important features are indeed important in terms of their contribution to prediction.

For quantifying robustness against perturbations, we adopt the metric proposed in 
\cite{zafar2021lack}, which is based on
the average Jaccard similarity between (i) the important features of an origi-nal example  
and (ii) the important features of the perturbed example  \cite{levandowsky1971distance}. 
The similarity is defined over interval $[0,1]$ such that a higher similarity indicates a more robust interpreter.

For quantifying effectiveness in coping with out-of-distribution examples, we adopt the metric defined in \cite{hooker2019benchmark}. 
Specifically, we take the number of features $n$ over 8, and incrementally and equally sample $q$ features among all the features, starting at $q=\frac{n}{8}$, i.e. $q\in Q=\{\frac{n}{8}, \frac{2n}{8}\cdots, \frac{7n}{8}\}$. 
For a given $q$, 
we use the same training set to learn the same kind of classifier $M_q$ by removing the
$q$ least important features (with respect to the interpreter), namely $\cup_{u_i\in U} \widetilde{E}(u_i)$, where $\widetilde{E}(u_i)$ is the code example $u_i$ with $q$ least important features (with respect to the interpreter) removed, and the difference of accuracy between classifier $M$ and retrained classifier $M_q$ is defined as $AD_q=\frac{|\{u_i\in U: M(u_i)=M(\widetilde{E}(u_i))\}|-|\{u_i\in U: M(u_i)=M_q(\widetilde{E}(u_i))\}|}{|U|}$.
The degree to which the interpreter is impacted by out-of-distribution inputs is the average $AD_q$ for each $q\in Q$. 
A smaller 
average difference of accuracy indicates a reduced impact of out-of-distribution inputs on the interpreter.

Corresponding to the preceding metrics, our experiments are driven by three {\em Research Questions} (RQs):

\begin{itemize}
\item[$\bullet$]\textbf{RQ1}: What is Robin's fidelity?  
(Section \ref{section:rq1})
\item[$\bullet$]\textbf{RQ2}: What is Robin's robustness against code perturbations? (Section \ref{section:rq2})
\item[$\bullet$]\textbf{RQ3}: What is Robin's effectiveness in coping with out-of-distribution examples? (Section \ref{section:rq3})
\end{itemize}

\subsection{Experimental Setup}

\smallskip
\noindent {\bf Implementation.}
We choose two deep learning-based code classifiers: DL-CAIS \cite{abuhamad2018large} for code authorship attribution and TBCNN \cite{mou2016convolutional} for code functionality classification.
We choose these two classifiers because they offer different code classification tasks, use different kinds of code representations and different neural networks, 
are representative of the state-of-the-art in code classification, and are open-sourced; these characteristics are necessary to test Robin's wide applicability. 

\begin{itemize}
\item \textbf{DL-CAIS} \cite{abuhamad2018large}. This classifier leverages a Term Frequency-Inverse Document Frequency based approach to extract lexical features from source code
and a {\em Recurrent Neural Network} (RNN) is employed to learn the code representation, which is then used as input to a random forest classifier to achieve code authorship attribution. In our experiment, we use a dataset from the {\em Google Code Jam} (GCJ) \cite{GCJ,DBLP:conf/uss/QuiringMR19}, involving 1,632 C++ program files from 204 authors for 8 programming challenges and has been widely used in code authorship attribution task \cite{DBLP:conf/uss/QuiringMR19, li2022ropgen}. This dataset is different from the one used in \cite{abuhamad2018large}, which is not available to us.

\item \textbf{TBCNN} \cite{mou2016convolutional}. The method represents source code as an {\em Abstract Syntax Tree} (AST), encodes the resulting AST as a vector, uses a tree-based convolutional layer to learn the features in the AST, and uses a fully-connected layer and softmax layer for making predictions.
In our experiment, we use the dataset of pedagogical programming Open Judge system, involving 52,000 C programs for 104 programming problems. This dataset is the same as the one used in \cite{mou2016convolutional} because it is publicly available.
\end{itemize}

We implement Robin in Python using Tensorflow \cite{abadi2016tensorflow} to retrofit the interpretability of DL-CAIS and TBCNN. We run experiments on a computer with 
a RTX A6000 GPU and an Intel Xeon Gold 6226R CPU operating at 2.90 GHz.

\smallskip
\noindent {\bf Interpreters for Comparison.}
We compare Robin with three existing interpreters: LIME \cite{ribeiro2016should}, LEMNA \cite{guo2018lemna}, and the one proposed in \cite{zou2021interpreting}, which would represent the state-of-the-art in interpretability of code classifier in feature-based  
post-hoc local interpretation. More specifically, LIME \cite{ribeiro2016should} makes small local perturbations to an example and obtains an interpretable linear regression model based on (i) the distance between the perturbed example and the original example and (ii) the change to the prediction. 
As such, LIME can be applied to explain any 
classifier.
LEMNA \cite{guo2018lemna} approximates local nonlinear decision boundaries for complex classifiers, especially RNN-based ones with sequential properties, to provide interpretations in security applications. Meanwhile, the method in \cite{zou2021interpreting} interprets vulnerability detector predictions by perturbing feature values, identifying important features based on their impact on predictions, training a decision-tree with the important features, and extracting rules for interpretation. Additionally, we establish a random feature selection method as a baseline.

\subsection{What Is Robin's Fidelity? (RQ1)}
\label{section:rq1}
To determine the effectiveness of Robin on fidelity, we first train two code classifiers DL-CAIS \cite{abuhamad2018large} and TBCNN \cite{mou2016convolutional} to be explained according to the settings of the literature, 
acheiving 88.24\% accuracy for code authorship attribution and 96.72\% accuracy for code functionality classification.
Then we apply Robin and the interpreters for comparison to DL-CAIS and TBCNN models. 
For Robin, we set the candidate number of selected coding style attributes $\theta_i$ to 4
and the number of important features selected by the interpreter $k$ to 10. We split the dataset randomly by 3:1:1 for training, validation, and testing for TBCNN and use 8-fold cross-validation for DL-CAIS when training the interpreter. 

\begin{table}[!tb]
\vspace{-0.2cm}
    \centering
    \footnotesize
    \caption{Fidelity evaluation results for different interpreters}
    \vspace{-0.2cm}
    \begin{tabular}{|c|c|c||c|c|}
        \hline
        \multirow{2}*{Method} & \multicolumn{2}{c||}{DL-CAIS} & \multicolumn{2}{c|}{TBCNN}\\
        \cline{2-5}
        ~ & FS-M (\%) & FS-A (\%) & FS-M (\%) & FS-A (\%) \\
        \hline
        Baseline & 1.96 & 2.45 & 10.29 & 9.23 \\ 
        \hline
        LIME \cite{ribeiro2016should}& 0.49 & 3.43 & 7.98 & 10.67 \\ 
        \hline
        LEMNA \cite{guo2018lemna}& 0.49 & 1.96 & 5.48 & 8.27 \\ 
        \hline
        Zou et al. \cite{zou2021interpreting} & {\bf 33.33} & 69.60 & 18.75 & 31.63 \\ 
        \hline
        Robin & 13.73 & {\bf 92.65} & {\bf 20.67} & {\bf 83.65} \\
        \hline
    \end{tabular}
    \label{tab:fidelity}
    \vspace{-0.4cm}
\end{table}

Table \ref{tab:fidelity} shows the fidelity evaluation results on DL-CAIS and TBCNN for different interpreters. We observe that LIME and LEMNA achieve an average FS-M of 0.49\% and an average FS-A of 2.70\% for DL-CAIS, and an average FS-M of 6.73\% and an average FS-A of 9.47\% for TBCNN, performing even worse than baseline.  
This can be explained by the fact that LIME and LEMNA do not perform well in multi-class code classification tasks due to the more complex decision boundaries of the classifiers.
We also observe that 
Robin outperforms other interpreters in terms of FS-M and FS-A metrics significantly except Zou et al.'s method \cite{zou2021interpreting} in terms of FS-M on DL-CAIS. Robin achieves 23.05\% higher FS-A at the cost of 19.60\% lower FS-M. However, Zou et al.'s method \cite{zou2021interpreting} is much less robust to perturbed examples than Robin which we will discuss in Section \ref{section:rq2}.
Compared with other interpreters, Robin achieves 6.11\% higher FS-M and 67.22\% higher FS-A on average, which indicates the high fidelity of Robin.

\begin{table}[!tb]
    \centering
    \footnotesize
    \caption{Average interpretation time of each code example for different interpreters}
    \vspace{-0.2cm}
    \begin{tabular}{|c|c|c|}
        \hline
        Method & DL-CAIS (ms) & TBCNN (ms) \\ \hline
        Baseline & {\bf 1.00} & {\bf 1.43} \\ \hline
        LIME \cite{ribeiro2016should} & 61957.71 & 111484.20 \\ \hline
        LEMNA \cite{guo2018lemna}& 17448.35 & 43722.97 \\ \hline
        Zou et al. \cite{zou2021interpreting} & 166298.43 & 243142.95 \\ \hline
        Robin & 1.71 & 408.04 \\ \hline
    \end{tabular}
    \label{tab:timecost}
    \vspace{-0.4cm}
\end{table}

For the time cost of interpreters, Table \ref{tab:timecost} shows the average interpretation time (in milliseconds) for each code example. We observe that Robin significantly outperforms the other three interpreters in terms of time cost. 
Note that while baseline is less time-consuming, it 
has much lower fidelity and robustness 
than Robin  
(see Section \ref{section:rq2}). 
Other interpreters are significantly more time costly than Robin 
because they are optimized independently on a single code example and require a new perturbation and analysis each time a code example is interpreted, while Robin directly constructs an interpreter that applies to all code examples and automatically identifies the important features by simply feeding code examples into the interpreter model. Robin achieves a 99.75\% reduction in time cost than the other three interpreters on average.

\smallskip
\noindent
{\bf Ablation Analysis.} 
Robin has two modules to improve
the interpreter, i.e., adding $L_{diff}$ to the loss of the interpreter (denoted as ``Factor1''), and data augmentation using mixup (denoted as ``Factor2''). 
To show the contribution of each module in Robin to the effectiveness of fidelity, we conduct the ablation study.
We exclude Factor1, Factor2, and both Factor1 and Factor2 to generate three variants of Robin, respectively, and compare Robin with the three variants in terms of fidelity.
Table \ref{tab:fidelity_ablation} summarizes the fidelity evaluation results of Robin and its variants on DL-CAIS and TBCNN.  
We observe that Robin without Factor1, Factor2, or both Factor1 and Factor2 can reduce FS-M of 1.48-1.97\% and FS-A of 1.96-4.90\% for DL-CAIS, and reduce FS-M of 0.48-1.73\% and FS-A of 1.53-2.88\% for TBCNN. Robin without Factor1 and Factor2 achieves the worst results.
This indicates the significance of Factor1 and Factor2 for the fidelity of Robin.

\begin{table}[!tb]
    \centering
    \footnotesize
    \caption{Ablation analysis results of fidelity evaluation (unit: \%)}
    \vspace{-0.2cm}
    \begin{tabular}{|c|c|c||c|c|}
        \hline
        \multirow{2}*{Method} & \multicolumn{2}{c||}{DL-CAIS} & \multicolumn{2}{c|}{TBCNN}\\
        \cline{2-5}
        ~ & FS-M & FS-A & FS-M & FS-A \\
        \hline
        Robin & {\bf 13.73} & {\bf 92.65} & {\bf 20.67} & {\bf 83.65} \\ 
        \hline
        Robin w/o Factor1 & 12.25 & 90.69 & 20.19 & 81.44 \\ 
        \hline
        Robin w/o Factor2 & 11.76 & 90.20 & 19.90 & 82.12 \\ 
        \hline
        Robin w/o Factor1\&2 & 12.25 & 87.75 & 18.94 & 80.77 \\ 
        \hline
    \end{tabular}
    \label{tab:fidelity_ablation}
    \vspace{-0.4cm}
\end{table}

\smallskip
\noindent
{\bf Effectiveness of Fidelity When Applied to Different Neural Network Structures.}
To demonstrate the applicability of Robin to various neural network structures, we take DL-CAIS for instance to replace the {\em Recurrent Neural Network} (RNN) layers of DL-CAIS with the {\em Convolutional Neural Network} (CNN) layers (denoted as ``DL-CAIS-CNN'') and replace the RNN layers of DL-CAIS with the {\em Multi-Layer Perception} (MLP) layers (denoted as ``DL-CAIS-MLP''), respectively. 
We first train two code authorship attribution models DL-CAIS-CNN and DL-CAIS-MLP to be explained according to the settings of DL-CAIS \cite{abuhamad2018large}. We obtain a DL-CAIS-CNN with an accuracy of 91.18\% and a DL-CAIS-MLP with an accuracy of 90.69\% for code authorship attribution.
Then we apply Robin and other interpreters for comparison to DL-CAIS-CNN and DL-CAIS-MLP respectively. 
Table
\ref{tab:fidelity_structure} shows the fidelity evaluation results for different interpreters on DL-CAIS with different neural networks. For DL-CAIS-CNN and DL-CAIS-MLP, Robin achieves a 40.07\% higher FS-M and an 83.50\% higher FS-A on average than the other three interpreters, which shows the effectiveness of Robin applied to different neural network structures.

\begin{table}[!tb]
    \centering
    \footnotesize
    \caption{Fidelity evaluation results for different interpreters on DL-CAIS with different neural network structures (unit: \%)}
    \vspace{-0.2cm}
        \begin{tabular}{|c|c|c||c|c||c|c|}
        \hline
        \multirow{2}*{Method} & \multicolumn{2}{c||}{DL-CAIS} & \multicolumn{2}{c||}{DL-CAIS-CNN} & \multicolumn{2}{c|}{DL-CAIS-MLP} \\
        \cline{2-7}
        ~ & FS-M & FS-A & FS-M & FS-A & FS-M & FS-A \\
        \hline
        Baseline & 1.96 & 2.45 & 2.94 & 3.92 & 4.41 & 3.43 \\ 
        \hline
        LIME \cite{ribeiro2016should}& 0.49 & 3.43 & 4.90 & 6.37 & 3.43 & 6.86 \\ 
        \hline
        LEMNA \cite{guo2018lemna}& 0.49 & 1.96 & 1.47 & 1.47 & 0.98 & 1.47 \\ 
        \hline
        Zou et al. \cite{zou2021interpreting} & {\bf 33.33} & 69.60 & 40.50 & 54.90 & 20.09 & 17.65\\ 
        \hline
        Robin & 13.73 & {\bf 92.65} & {\bf 40.69} & {\bf 99.51} & {\bf 63.24} & {\bf 97.06}\\
        \hline
    \end{tabular}
    \label{tab:fidelity_structure}
    \vspace{-0.4cm}
\end{table}

\begin{figure*}[!tb]
  \centering
  \includegraphics[width=0.95\linewidth]{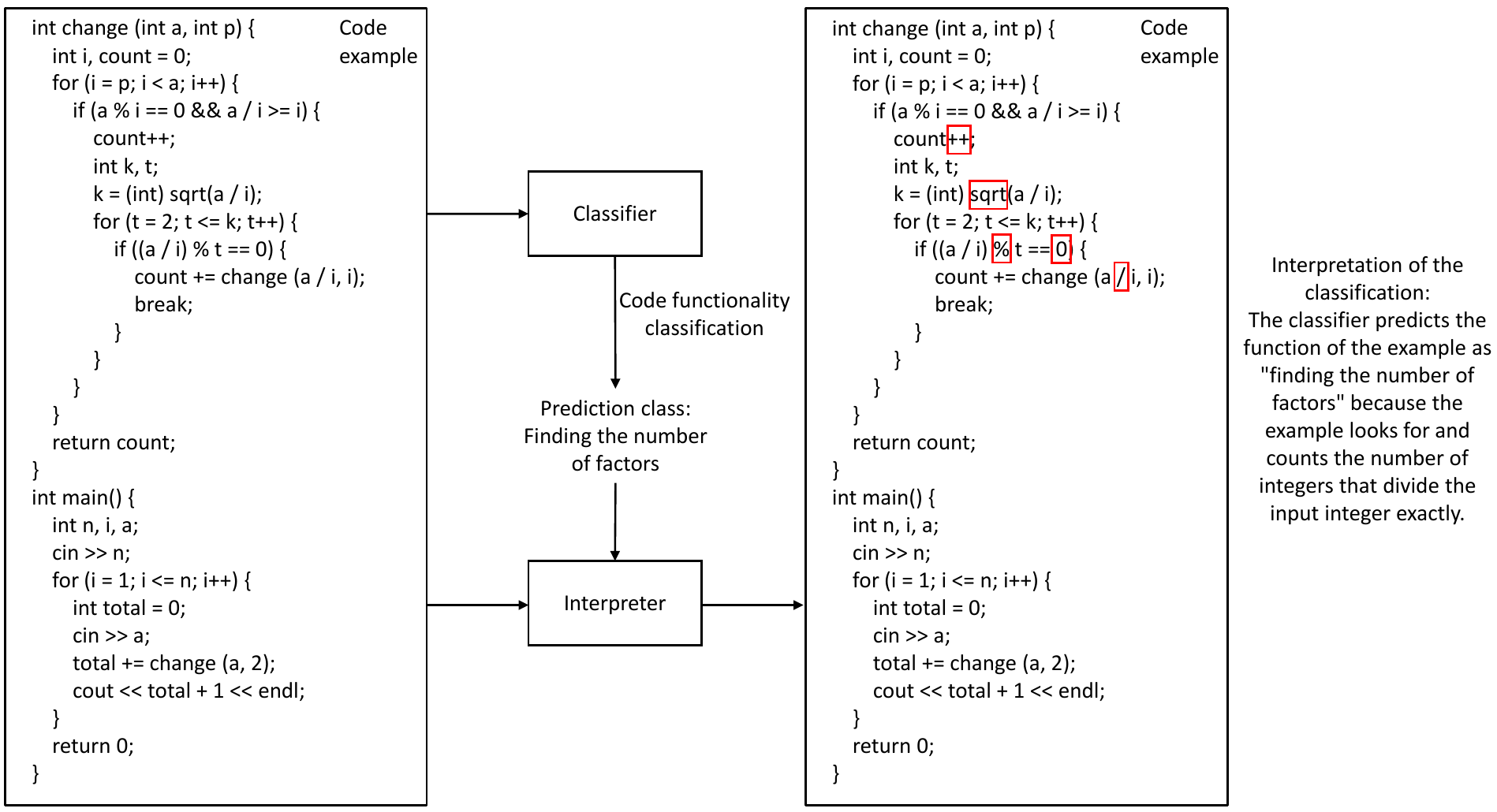}
  \vspace{-0.2cm}
  \caption{The interpretation of a specific instance of code classification in the context of code functionality classification, where red boxes highlight the 5 most important features.}
  \label{fig:case}
  \vspace{-0.5cm}
\end{figure*}

\smallskip
\noindent
{\bf Usefulness of Robin in Understanding Reasons for Classification.}
To illustrate the usefulness of Robin in this perspective, we consider a scenario of code functionality classification via TBCNN \cite{mou2016convolutional}. The code example in Fig. \ref{fig:case} is predicted by the classifier as the functionality class ``finding the number of factors''. The interpreter generated by Robin extracts five features of the code example, which are deemed most relevant with respect to the prediction result and are highlighted via red boxes in Fig. \ref{fig:case}. These five features are related to the remainder, division, and counting operators. By analyzing these five features, it becomes clear that 
the code example is predicted as ``finding the number of factors'' because the example looks for, and counts, the number of integers that can divide the input integer.

\begin{insight}
{\em Robin achieves a 6.11\% higher FS-M and a 67.22\% higher FS-A on average than 
the three interpreters we considered.}
\end{insight}

\subsection{What Is Robin's Robustness? (RQ2)}
\label{section:rq2}
To evaluate the robustness of Robin against perturbations, we generate perturbed examples by using the semantics-preserving
code transformation to code examples in the test set 
and filter out the perturbed examples that change
the predicted labels of the classifier. 
We use these perturbed examples to test the robustness of interpreters.

\begin{table}[!tb]
    \centering
    \footnotesize
    \caption{Robustness evaluation results for different interpreters}
    \vspace{-0.2cm}
    \begin{tabular}{|c|c|c|}
        \hline
        Method & DL-CAIS & TBCNN\\ \hline
        Baseline & 0.0121 & 0.0348 \\ \hline
        LIME \cite{ribeiro2016should} &  0.0592 & 0.0962 \\ \hline
        LEMNA \cite{guo2018lemna}& 0.0157 & 0.0475 \\ \hline
        Zou et al. \cite{zou2021interpreting} & 0.3681 & 0.3852 \\ \hline
        Robin & {\bf 0.9275} & {\bf 0.5269} \\ \hline
    \end{tabular}
    \label{tab:robustness}
    \vspace{-0.6cm}
\end{table}

Table \ref{tab:robustness} summarizes the robustness evaluation results for different interpreters. 
We observe that the robustness of LIME and LEMNA on the code classifier is very poor and only slightly higher than the baseline. This is caused by the following: LIME and LEMNA suffer from uncertainty, thus there may be differences between the important features obtained when the same code example is interpreted multiple times. 
We also observe that the robustness of Zou et al. 's method \cite{zou2021interpreting} is higher than that of LIME and LEMNA, but still much lower than 
that of Robin. 
The average Jaccard similarity between the important features of the original examples identified by Robin and the important features of the adversarial examples is 1.94x 
higher than the state-of-the-art method \cite{zou2021interpreting} and 15.87x 
higher on average than the three interpreters we evaluated for DL-CAIS and TBCNN. This indicates that Robin is insensitive to semantics-preserving code transformations
and has higher robustness against perturbations.

\smallskip
\noindent
{\bf Ablation Analysis.} 
To show the contribution of Factor1 and Factor2 in Robin to the robustness, we conduct the ablation study.
We exclude Factor1, Factor2, and both Factor1 and Factor2 to generate three variants of Robin, respectively, and compare Robin with the three variants in terms of robustness for the number of important features $k=10$. 
Table \ref{tab:robustness_ablation} summarizes the robustness evaluation results of Robin and its three variants on DL-CAIS and TBCNN. 
We observe that Robin achieves the highest robustness, and removing Factor1 and/or Factor2 can decrease its robustness, which indicates the significance of Factor1 and Factor2 to Robin's robustness.

\begin{table}[!tb]
    \centering
    \footnotesize
    \caption{Ablation analysis results of robustness evaluation
    }
    \vspace{-0.2cm}
    \begin{tabular}{|c|c|c|}
        \hline
        Method & DL-CAIS & TBCNN \\ 
        \hline
        Robin & {\bf 0.9275} & {\bf 0.5269} \\ \hline
        Robin w/o Factor1 & 0.9181 & 0.5194 \\ \hline
        Robin w/o Factor2 & 0.9174 & 0.5025 \\ \hline
        Robin w/o Factor1\&2 & 0.9073 & 0.4931 \\ \hline
    \end{tabular}
    \label{tab:robustness_ablation}
    \vspace{-0.4cm}
\end{table}

To show the impact of the number of important features $k$ on the robustness, we take DL-CAIS for example to compare Robin and its three variants when applied to DL-CAIS in terms of the robustness of interpreters 
based on $k$ (e.g., 10, 20, 30, 40, and 50) important features, respectively.
As shown in Table \ref{tab:robustness_ablation_k},  
the robustness decreases as $k$ increases. This can be explained by the following: As $k$ increases, the less important features
are added to the selected important features; 
these less important features are difficult to be recognized by the interpreter due to their less prominent contribution to the prediction,  
thus perform worse robustness against perturbations. We also observe that (i) Robin achieves the best robustness on DL-CAIS in all $k$ values, and (ii) removing Factor1 or Factor2 or both of them from Robin can decrease the robustness of Robin, which indicates the significance of Factor1 and Factor2 for the robustness of Robin.

\begin{table}[!tb]
    \centering
    \footnotesize
    \caption{Ablation analysis results of robustness evaluation on DL-CAIS in different $k$ value
    }
    \vspace{-0.2cm}
    \begin{tabular}{|c|c|c|c|c|c|}
        \hline
        Method & $k$=10 & $k$=20 & $k$=30 & $k$=40 & $k$=50\\ 
        \hline
        Robin & {\bf 0.9275} & {\bf 0.9129} & {\bf 0.8962} & {\bf 0.8939} & {\bf 0.8835} \\ \hline
        Robin w/o Factor1 & 0.9181 & 0.8981 & 0.8932 & 0.8819 & 0.8808 \\ \hline
        Robin w/o Factor2 & 0.9174 & 0.8969 & 0.8793 & 0.8799 & 0.8784 \\ \hline
        Robin w/o Factor1\&2 & 0.9073 & 0.8943 & 0.8744 & 0.8731 & 0.8640 \\ \hline
    \end{tabular}
    \label{tab:robustness_ablation_k}
    \vspace{-0.4cm}
\end{table}

\smallskip
\noindent
{\bf Robustness Evaluation When Applied to Different Neural Network Structures.}
To show the robustness of Robin when applied to different neural network structures, we 
adopt DL-CAIS, DL-CAIS-CNN, and DL-CAIS-MLP we have trained in Section \ref{section:rq1} for interpretation.
For DL-CAIS-CNN and DL-CAIS-MLP, we generate perturbed examples by using the semantics-preserving code transformations to code examples in the test set and filter out the perturbed examples that change the prediction labels of the classifier. 
Table \ref{tab:robustness_structure} shows the robustness evaluation results for different interpreters on DL-CAIS with different neural networks. For DL-CAIS-CNN and DL-CAIS-MLP, Robin achieves a 10.05x higher robustness on average, compared with the other three interpreters.
Though Robin achieves different robustness for different neural network structures, Robin achieves the highest robustness among all interpreters we evaluated.

\begin{table}[!tb]
    \centering
    \footnotesize
    \caption{Robustness evaluation results of different interpreters on DL-CAIS with different neural network structures}
    \vspace{-0.2cm}
    \begin{tabular}{|c|c|c|c|}
        \hline
        Method & DL-CAIS & DL-CAIS-CNN & DL-CAIS-MLP \\ \hline
        Baseline & 0.0121 & 0.0121 & 0.0121 \\ \hline
        LIME \cite{ribeiro2016should} & 0.0592 & 0.2651 & 0.0592 \\ \hline
        LEMNA \cite{guo2018lemna} & 0.0157 & 0.0167 & 0.0157 \\ \hline
        Zou et al. \cite{zou2021interpreting} & 0.3681 & 0.3812 & 0.3059 \\ \hline
        Robin & {\bf 0.9275} & {\bf 0.4922} & {\bf 0.3298} \\ \hline
    \end{tabular}
    \label{tab:robustness_structure}
    \vspace{-0.4cm}
\end{table}

\begin{insight}
{\em Robin achieves a 1.94x higher robustness than the state-of-the-art method \cite{zou2021interpreting} and a 15.87x higher robustness on average than 
the three interpreters we evaluated.}
\end{insight}

\subsection{What Is Robin's Effectiveness in Coping with Out-of-Distribution Examples? (RQ3)}
\label{section:rq3}
To demonstrate the effectiveness of Robin in copying with out-of-distribution examples, we conduct experiments with the number of removed non-important features $q\in Q$=\{100, 200, 300, 400, 500, 600, 700\} for DL-CAIS and $q\in Q$=\{25, 50, 75, 100, 125, 150, 175\} for TBCNN according to the number of all features. 
Table \ref{tab:ood} shows the difference of accuracy $AD_q$ 
between the classifier and the retrained classifier with $q$ non-important features removed. We observe that the average difference of accuracy of Robin and the baseline method is very small,  
indicating that they are less affected by out-of-distribution examples. This can be explained by the fact that both of these methods do not employ the change in classifier's accuracy to assess the importance of features. Although the baseline method outperforms Robin on DL-CAIS, it has much lower fidelity and robustness than Robin which we have discussed in Section \ref{section:rq1} and Section \ref{section:rq2}. In contrast, the average difference of accuracy achieved by LEMNA is notably 
larger
than those of Robin and the baseline method, 
because LEMNA relies on the changes of classifier's accuracy to calculate the importance of features. 
Robin achieves a 24.21\% smaller average difference of accuracy for DL-CAIS and a 70.41\% smaller average difference of accuracy for TBCNN than LEMNA, indicating that Robin achieves 47.31\% less affected by the out-of-distribution examples compared to LEMNA on average. 
Robin is minimally affected by the out-of-distribution examples, which attributes to introducing the prediction accuracy of the retrained classifier to evaluate the importance of features.  
 
\begin{table}[!tb]
    \centering
    \tiny
    \caption{The difference of accuracy $AD_q$ between classifier and retrained classifier with $q$ non-important features removed}
    \vspace{-0.2cm}
    \begin{tabular}{|c|c|c|c|c|c|c|c|c|}
        \hline
        \multicolumn{9}{|c|}{DL-CAIS}
        \\ \hline
        Method & $q$=100 & $q$=200 & $q$=300 & $q$=400 & $q$=500 & $q$=600 & $q$=700 & Average \\ \hline
        Baseline & 0.1029 & 0.1029 & 0.1274 & 0.0979 & 0.098 & 0.0735 & 0.0931 & {\bf 0.0994}
        \\ \hline
        LEMNA & 0.0834 & 0.0784 & 0.1030 & 0.0637 & 0.3579 & 0.2402 & 0.0245 & 0.1359
        \\ \hline
        Robin & 0.0736 & 0.0834 & 0.0785 & 0.0883 & 0.1030 & 0.1128 & 0.1814 & 0.1030
        \\ \hline
        \hline
        \multicolumn{9}{|c|}{TBCNN}
        \\ \hline
        Method & $q$=25 & $q$=50 & $q$=75 & $q$=100 & $q$=125 & $q$=150 & $q$=175 & Average
        \\ \hline
        Baseline & 0.0074 & 0.0042 & 0.0204 & 0.0138 & 0.0043 & 0.0355 & 0.0358 & 0.0173
        \\ \hline
        LEMNA & 0.0105 & 0.0159 & 0.0241 & 0.0227 & 0.1804 & 0.0569 & 0.0445 & 0.0507
        \\ \hline
        Robin & 0.0299 & 0.0037 & 0.0184 & 0.0156 & 0.0200 & 0.0029 & 0.0145 & {\bf 0.0150}       
        \\ \hline
    \end{tabular}
    \label{tab:ood}
    \vspace{-0.4cm}
\end{table}

\smallskip
\noindent
{\bf Effectiveness in Coping with Out-of-Distribution Examples When Applied to Different Neural Network Structures.} To show the effectiveness in coping with out-of-distribution examples when applied to different neural network structures, we adopt DL-CAIS, DL-CAIS-CNN, and DL-CAIS-MLP we have trained in Section \ref{section:rq1} for interpretation.
Table \ref{tab:ood_structure} describes the difference of accuracy $AD_q$ between the given classifier and the retrained classifier after removing $q$ non-important features while using 
different neural network structures. 
For DL-CAIS-CNN and DL-CAIS-MLP, Robin achieves a 70.00\% less affected by out-of-distribution examples when compared with LEMNA on average, where the average is taken over DL-CAIS-CNN and DL-CAIS-MLP, which shows the effectiveness of Robin in coping with out-of-distribution examples when applied to different neural network structures.

\begin{table}[!tb]
    \centering
    \tiny
    \caption{The difference of accuracy $AD_q$ between classifier and retrained classifier with $q$ non-important features removed on DL-CAIS with different neural network structures}
    \vspace{-0.2cm}
    \begin{tabular}{|c|c|c|c|c|c|c|c|c|}
        \hline
        \multicolumn{9}{|c|}{DL-CAIS}
        \\ \hline
        Method & $q$=100 & $q$=200 & $q$=300 & $q$=400 & $q$=500 & $q$=600 & $q$=700 & Average \\ \hline
        Baseline & 0.1029 & 0.1029 & 0.1274 & 0.0979 & 0.098 & 0.0735 & 0.0931 & {\bf 0.0994}
        \\ \hline
        LEMNA & 0.0834 & 0.0784 & 0.1030 & 0.0637 & 0.3579 & 0.2402 & 0.0245 & 0.1359
        \\ \hline
        Robin & 0.0736 & 0.0834 & 0.0785 & 0.0883 & 0.1030 & 0.1128 & 0.1814 & 0.1030
        \\ \hline
        \hline
        \multicolumn{9}{|c|}{DL-CAIS-CNN}
        \\ \hline
        Method & $q$=100 & $q$=200 & $q$=300 & $q$=400 & $q$=500 & $q$=600 & $q$=700 & Average \\ \hline
        Baseline & 0.0147 & 0.0049 & 0.0147 & 0.0049 & 0.0049 & 0.0196 & 0.0818 & {\bf 0.0207}
        \\ \hline
        LEMNA & 0.0490 & 0.0396 & 0.0245 & 0.0195 & 0.2745 & 0.1618 & 0.0196 & 0.0840
        \\ \hline
        Robin & 0.0196 & 0.0049 & 0.0095 & 0.0196 & 0.0294 & 0.0294 & 0.0735 & 0.0266
        \\ \hline
        \hline
        \multicolumn{9}{|c|}{DL-CAIS-MLP}
        \\ \hline
        Method & $q$=100 & $q$=200 & $q$=300 & $q$=400 & $q$=500 & $q$=600 & $q$=700 & Average 
        \\ \hline
        Baseline & 0.0197 & 0.0196 & 0.0588 & 0.0147 & 0.0049 & 0.0687 & 0.0765 & 0.0376
        \\ \hline
        LEMNA & 0.0049 & 0.0047 & 0.0196 & 0.0147 & 0.2598 & 0.2010 & 0.0490 & 0.0791
        \\ \hline
        Robin & 0.0098 & 0.0049 & 0.0049 & 0.0196 & 0.0196 & 0.0490 & 0.0490 & {\bf 0.0224}    
        \\ \hline
    \end{tabular}
    \label{tab:ood_structure}
    \vspace{-0.6cm}
\end{table}

\begin{insight}
{\em Robin achieves a 
47.31\% less affected by out-of-distribution examples when compared with LEMNA.}
\end{insight}

\section{Limitations}
\label{sec:limitations}
The present study has limitations, which represent exciting open problems for future studies. 
First, our study does not evaluate the effectiveness of Robin on graph-based code classifiers and pre-training models like CodeT5\cite{wang-etal-2021-codet5} and CodeBERT\cite{feng-etal-2020-codebert}. The unique characteristics of these models pose challenges that require further investigation, particularly in the context of applying Robin to classifiers with more complex model structures.
Second, Robin can identify the most important features but cannot give further explanations why a particular prediction is made. To our knowledge, this kind of desired further explanation is beyond the reach of the current technology in deep learning interpretability. 
Third, Robin can identify the most important features that lead to the particular prediction of a given example, but cannot tell which training examples in the training set that leads to the code classifier contribute to the particular prediction. Achieving this type of training examples traceability is important because it may help achieve better interpretability.

\section{Related Work}
\label{sec:related_work}

\noindent
{\bf Prior Studies on Deep Learning-Based Code Classifiers.}
We divide these models into three categories according to the code representation they use: 
{\em token-based} \cite{russell2018automated, abuhamad2018large, yang2017authorship} vs. {\em tree-based} \cite{alon2019code2vec, lin2017poster, alsulami2017source} vs. {\em graph-based} \cite{allamanis2017learning, li2021sysevr, zou2022mvulpreter}.
Token-based models represent a piece of code as a sequence of individual tokens, while only performing basic lexical analysis. These models are mainly used for code authorship attribution and 
vulnerability detection. Tree-based models represent a piece of code as a syntax tree, while incorporating both lexical and syntax analysis. These models are widely used for code authorship attribution, code function classification, and vulnerability detection. Graph-based models represent a piece of code as a directed graph, where a node represents an expression or statement and an edge represents a control flow, control dependence, or data dependence. These models are suitable for complex code structures such as vulnerability detection.
We have shown how Robin can offer interpretability to token- and tree-based code classifiers \cite{abuhamad2018large,mou2016convolutional}, 
but not to graph-based models as discussed in the preceding section.

\smallskip
\noindent{\bf Prior Studies on Interpretation Methods for Deep Learning Models.}
These studies are often divided into two approaches: {\em ante-hoc} \cite{vaswani2017attention, choi2016retain} vs. {\em post-hoc} \cite{puri2017magix, wang2019deepvid, simonyan2013deep, ribeiro2016should, lundberg2017unified, schwab2019cxplain, smilkov2017smoothgrad}, where the latter can be further divided into {\em global} (i.e., seeking model-level interpretability) \cite{puri2017magix, wang2019deepvid} vs. {\em local} (i.e., seeking example-level interpretability) \cite{simonyan2013deep, ribeiro2016should, lundberg2017unified, schwab2019cxplain, smilkov2017smoothgrad} interpretation methods. 
In the context of code classification, the ante-hoc approach leverages the attention weight matrix \cite{bui2019autofocus, zou2022mvulpreter}. There is currently no post-hoc approach aiming at global interpretation in code classification;
whereas, the post-hoc approach aiming at local interpretation
mainly leverages perturbation-based feature saliency \cite{zou2021interpreting, cito2022counterfactual} and program reduction \cite{suneja2021probing, rabin2021understanding}.
Since ante-hoc interpretation methods cannot provide interpretations for given classifiers, we will not discuss them any further. On the other hand, existing poc-hoc methods  
are not robust (Section \ref{section:rq2}); in particular, existing methods for local interpretation suffers from the problem of out-of-distribution examples \cite{hooker2019benchmark, brocki2022evaluation}. Robin addresses both the robustness issue and the out-of-distribution issue in the post-hoc approach to local interpretation, 
by introducing approximators to mitigate out-of-distribution examples and using adversarial training and data augmentation to improve robustness.

\smallskip
\noindent
{\bf Prior Studies on Improving Robustness of Interpretation Methods.}
These studies have been conducted in other application domains than code classification.
In the image domain, one idea is to aggregate multiple interpretation \cite{rieger2020simple, zhang2020interpretable}, and another idea is to 
smooth the model's decision surface \cite{wang2020smoothed, smilkov2017smoothgrad}. 
In the text domain, one idea is to eliminate the uncertainties that are present in the existing interpretation methods \cite{zhao2021baylime, zhou2021s}, and another idea is to introduce continuous small perturbations to interpretation and use adversarial training for robustness enhancement \cite{lakkaraju2020robust, la2021guaranteed}.
To our knowledge, we are the first to investigate how to achieve robust interpretability in the code classification domain, while noting that
none of the aforementioned methods that are effective in the other domains can be adapted 
to the code classification domain. This is because program code must follow strict lexical and syntactic requirements, meaning that perturbed representations  
may not be mapped back to real-world code examples, which is a general challenge when dealing with programs.
This justifies why Robin initiates the study of a new and important problem.

\section{Conclusion}
\label{sec:conclusion}

We have presented Robin, a robust interpreter for deep learning-based 
code classifiers such as code authorship attribution classification and code function classification.
The key idea behind Robin is to (i) use approximators to mitigate the out-of-distribution example problem, and (ii) use adversarial training and data augmentation to improve interpreter robustness, which is different from the widely-adopted idea of using adversarial training to achieve classifier's (rather than interpreter's) robustness. Experimental results show that Robin achieves a high fidelity and a high robustness, while mitigating the effect of out-of-distribution examples caused by perturbations. 
The limitations of Robin serve as interesting open problems for future research.

\section*{Acknowledgments}
We thank the anonymous reviewers for their comments which guided us in improving the paper. The authors affiliated with Huazhong University of Science and Technology were supported by the National Natural Science Foundation of China under Grant No. 62272187.
Shouhuai Xu was supported in part by the National Science Foundation under Grants \#2122631, \#2115134, and \#1910488 as well as Colorado State Bill 18-086. 
Any opinions, findings, conclusions or recommendations expressed in this work are those of the authors and do not reflect the views of the funding agencies in any sense.

\bibliographystyle{IEEEtran}
\bibliography{reference}

\begin{thebibliography}{10}
\providecommand{\url}[1]{#1}
\csname url@samestyle\endcsname
\providecommand{\newblock}{\relax}
\providecommand{\bibinfo}[2]{#2}
\providecommand{\BIBentrySTDinterwordspacing}{\spaceskip=0pt\relax}
\providecommand{\BIBentryALTinterwordstretchfactor}{4}
\providecommand{\BIBentryALTinterwordspacing}{\spaceskip=\fontdimen2\font plus
\BIBentryALTinterwordstretchfactor\fontdimen3\font minus
  \fontdimen4\font\relax}
\providecommand{\BIBforeignlanguage}[2]{{%
\expandafter\ifx\csname l@#1\endcsname\relax
\typeout{** WARNING: IEEEtran.bst: No hyphenation pattern has been}%
\typeout{** loaded for the language `#1'. Using the pattern for}%
\typeout{** the default language instead.}%
\else
\language=\csname l@#1\endcsname
\fi
#2}}
\providecommand{\BIBdecl}{\relax}
\BIBdecl

\bibitem{zhang2019novel}
J.~Zhang, X.~Wang, H.~Zhang, H.~Sun, K.~Wang, and X.~Liu, ``A novel neural
  source code representation based on abstract syntax tree,'' in
  \emph{Proceedings of the 41st International Conference on Software
  Engineering (ICSE), QC, Canada}.\hskip 1em plus 0.5em minus 0.4em\relax IEEE,
  2019, pp. 783--794.

\bibitem{mou2016convolutional}
L.~Mou, G.~Li, L.~Zhang, T.~Wang, and Z.~Jin, ``Convolutional neural networks
  over tree structures for programming language processing,'' in
  \emph{Proceedings of the 30th AAAI Conference on Artificial Intelligence
  (AAAI), Phoenix, Arizona, USA}.\hskip 1em plus 0.5em minus 0.4em\relax AAAI
  Press, 2016, pp. 1287--1293.

\bibitem{caliskan2015anonymizing}
A.~Caliskan-Islam, R.~Harang, A.~Liu, A.~Narayanan, C.~Voss, F.~Yamaguchi, and
  R.~Greenstadt, ``De-anonymizing programmers via code stylometry,'' in
  \emph{Proceedings of the 24th USENIX Security Symposium (USENIX Security),
  Washington, D.C., USA}, 2015, pp. 255--270.

\bibitem{alsulami2017source}
B.~Alsulami, E.~Dauber, R.~Harang, S.~Mancoridis, and R.~Greenstadt, ``Source
  code authorship attribution using long short-term memory based networks,'' in
  \emph{Proceedings of the 22nd European Symposium on Research in Computer
  Security (ESORICS), Oslo, Norway}, 2017, pp. 65--82.

\bibitem{yang2017authorship}
X.~Yang, G.~Xu, Q.~Li, Y.~Guo, and M.~Zhang, ``Authorship attribution of source
  code by using back propagation neural network based on particle swarm
  optimization,'' \emph{PloS one}, vol.~12, no.~11, p. e0187204, 2017.

\bibitem{bogomolov2021authorship}
E.~Bogomolov, V.~Kovalenko, Y.~Rebryk, A.~Bacchelli, and T.~Bryksin,
  ``Authorship attribution of source code: A language-agnostic approach and
  applicability in software engineering,'' in \emph{Proceedings of the 29th ACM
  Joint Meeting on European Software Engineering Conference and Symposium on
  the Foundations of Software Engineering (ESEC/FSE), Athens, Greece}, 2021,
  pp. 932--944.

\bibitem{abuhamad2018large}
M.~Abuhamad, T.~AbuHmed, A.~Mohaisen, and D.~Nyang, ``Large-scale and
  language-oblivious code authorship identification,'' in \emph{Proceedings of
  the 2018 ACM SIGSAC Conference on Computer and Communications Security (CCS),
  Toronto, ON, Canada}, 2018, pp. 101--114.

\bibitem{lin2017poster}
G.~Lin, J.~Zhang, W.~Luo, L.~Pan, and Y.~Xiang, ``Poster: Vulnerability
  discovery with function representation learning from unlabeled projects,'' in
  \emph{Proceedings of the 2017 ACM SIGSAC Conference on Computer and
  Communications Security (CCS), Dallas, TX, USA}, 2017, pp. 2539--2541.

\bibitem{li2018vuldeepecker}
Z.~Li, D.~Zou, S.~Xu, X.~Ou, H.~Jin, S.~Wang, Z.~Deng, and Y.~Zhong,
  ``{VulDeePecker}: A deep learning-based system for vulnerability detection,''
  in \emph{Proceedings of the 25th Annual Network and Distributed System
  Security Symposium (NDSS), San Diego, California, USA}, 2018, pp. 1--15.

\bibitem{li2021sysevr}
Z.~Li, D.~Zou, S.~Xu, H.~Jin, Y.~Zhu, and Z.~Chen, ``{SySeVR}: A framework for
  using deep learning to detect software vulnerabilities,'' \emph{IEEE
  Transactions on Dependable and Secure Computing}, vol.~19, no.~4, pp.
  2244--2258, 2022.

\bibitem{vaswani2017attention}
A.~Vaswani, N.~Shazeer, N.~Parmar, J.~Uszkoreit, L.~Jones, A.~N. Gomez,
  {\L}.~Kaiser, and I.~Polosukhin, ``Attention is all you need,'' in
  \emph{Proceedings of Annual Conference on Neural Information Processing
  Systems (NeurIPS), Long Beach, CA, USA}, 2017, pp. 5998--6008.

\bibitem{choi2016retain}
E.~Choi, M.~T. Bahadori, J.~Sun, J.~Kulas, A.~Schuetz, and W.~Stewart,
  ``Retain: An interpretable predictive model for healthcare using reverse time
  attention mechanism,'' in \emph{Proceedings of Annual Conference on Neural
  Information Processing Systems (NeurIPS), Barcelona, Spain}, 2016, pp.
  3504--3512.

\bibitem{ribeiro2016should}
M.~T. Ribeiro, S.~Singh, and C.~Guestrin, ``{`Why should I trust you?'
  E}xplaining the predictions of any classifier,'' in \emph{Proceedings of the
  22nd ACM SIGKDD International Conference on Knowledge Discovery and Data
  Mining (KDD), San Francisco, CA, USA}, 2016, pp. 1135--1144.

\bibitem{bui2019autofocus}
N.~D. Bui, Y.~Yu, and L.~Jiang, ``Autofocus: Interpreting attention-based
  neural networks by code perturbation,'' in \emph{Proceedings of the 34th
  IEEE/ACM International Conference on Automated Software Engineering (ASE),
  San Diego, CA, USA}, 2019, pp. 38--41.

\bibitem{zou2022mvulpreter}
D.~Zou, Y.~Hu, W.~Li, Y.~Wu, H.~Zhao, and H.~Jin, ``{mVulPreter}: A
  multi-granularity vulnerability detection system with interpretations,''
  \emph{IEEE Transactions on Dependable and Secure Computing}, pp. 1--12, 2022.

\bibitem{zou2021interpreting}
D.~Zou, Y.~Zhu, S.~Xu, Z.~Li, H.~Jin, and H.~Ye, ``Interpreting deep
  learning-based vulnerability detector predictions based on heuristic
  searching,'' \emph{ACM Transactions on Software Engineering and Methodology},
  vol.~30, no.~2, pp. 1--31, 2021.

\bibitem{cito2022counterfactual}
J.~Cito, I.~Dillig, V.~Murali, and S.~Chandra, ``Counterfactual explanations
  for models of code,'' in \emph{Proceedings of the 44th IEEE/ACM International
  Conference on Software Engineering: Software Engineering in Practice
  (ICSE-SEIP), Pittsburgh, PA, USA}, 2022, pp. 125--134.

\bibitem{suneja2021probing}
S.~Suneja, Y.~Zheng, Y.~Zhuang, J.~A. Laredo, and A.~Morari, ``Probing model
  signal-awareness via prediction-preserving input minimization,'' in
  \emph{Proceedings of the 29th ACM Joint Meeting on European Software
  Engineering Conference and Symposium on the Foundations of Software
  Engineering (ESEC/FSE), Athens, Greece}, 2021, pp. 945--955.

\bibitem{rabin2021understanding}
M.~R.~I. Rabin, V.~J. Hellendoorn, and M.~A. Alipour, ``Understanding neural
  code intelligence through program simplification,'' in \emph{Proceedings of
  the 29th ACM Joint Meeting on European Software Engineering Conference and
  Symposium on the Foundations of Software Engineering (ESEC/FSE), Athens,
  Greece}, 2021, pp. 441--452.

\bibitem{zeller2002simplifying}
A.~Zeller and R.~Hildebrandt, ``Simplifying and isolating failure-inducing
  input,'' \emph{IEEE Transactions on Software Engineering}, vol.~28, no.~2,
  pp. 183--200, 2002.

\bibitem{hooker2019benchmark}
S.~Hooker, D.~Erhan, P.-J. Kindermans, and B.~Kim, ``A benchmark for
  interpretability methods in deep neural networks,'' in \emph{Proceedings of
  Annual Conference on Neural Information Processing Systems (NeurIPS),
  Vancouver, BC, Canada}, 2019, pp. 9734--9745.

\bibitem{brocki2022evaluation}
L.~Brocki and N.~C. Chung, ``Evaluation of interpretability methods and
  perturbation artifacts in deep neural networks,'' \emph{arXiv preprint
  arXiv:2203.02928}, 2022.

\bibitem{bajaj2021robust}
M.~Bajaj, L.~Chu, Z.~Y. Xue, J.~Pei, L.~Wang, P.~C.-H. Lam, and Y.~Zhang,
  ``Robust counterfactual explanations on graph neural networks,'' in
  \emph{Proceedings of Annual Conference on Neural Information Processing
  Systems (NeurIPS), Virtual Event}, 2021, pp. 5644--5655.

\bibitem{zhang2020interpretable}
X.~Zhang, N.~Wang, H.~Shen, S.~Ji, X.~Luo, and T.~Wang, ``Interpretable deep
  learning under fire,'' in \emph{Proceedings of the 29th USENIX Security
  Symposium (USENIX Security), Virtual Event}, 2020, pp. 1659--1676.

\bibitem{guo2018lemna}
W.~Guo, D.~Mu, J.~Xu, P.~Su, G.~Wang, and X.~Xing, ``{LEMNA}: Explaining deep
  learning based security applications,'' in \emph{Proceedings of the 2018 ACM
  SIGSAC Conference on Computer and Communications Security (CCS), Toronto, ON,
  Canada}, 2018, pp. 364--379.

\bibitem{chen2018learning}
J.~Chen, L.~Song, M.~Wainwright, and M.~Jordan, ``Learning to explain: An
  information-theoretic perspective on model interpretation,'' in
  \emph{Proceedings of the 35th International Conference on Machine Learning
  (ICML), Stockholmsm{\"{a}}ssan, Stockholm, Sweden}, 2018, pp. 883--892.

\bibitem{lakkaraju2020robust}
H.~Lakkaraju, N.~Arsov, and O.~Bastani, ``Robust and stable black box
  explanations,'' in \emph{Proceedings of the 37th International Conference on
  Machine Learning (ICML), Virtual Event}, 2020, pp. 5628--5638.

\bibitem{la2021guaranteed}
E.~La~Malfa, A.~Zbrzezny, R.~Michelmore, N.~Paoletti, and M.~Kwiatkowska, ``On
  guaranteed optimal robust explanations for {NLP} models,'' in
  \emph{Proceedings of the 30th International Joint Conference on Artificial
  Intelligence (IJCAI), Virtual Event}, 2021, pp. 2658--2665.

\bibitem{zhang2017mixup}
H.~Zhang, M.~Cisse, Y.~N. Dauphin, and D.~Lopez-Paz, ``mixup: Beyond empirical
  risk minimization,'' in \emph{Proceedings of the 6th International Conference
  on Learning Representations (ICLR), Vancouver, BC, Canada}, 2018.

\bibitem{li2022ropgen}
Z.~Li, G.~Chen, C.~Chen, Y.~Zou, and S.~Xu, ``{RopGen}: Towards robust code
  authorship attribution via automatic coding style transformation,'' in
  \emph{Proceedings of the 44th International Conference on Software
  Engineering (ICSE), Pittsburgh, PA, USA}, 2022, pp. 1906--1918.

\bibitem{levandowsky1971distance}
M.~Levandowsky and D.~Winter, ``Distance between sets,'' \emph{Nature}, vol.
  234, no. 5323, pp. 34--35, 1971.

\bibitem{liang2020adversarial}
J.~Liang, B.~Bai, Y.~Cao, K.~Bai, and F.~Wang, ``Adversarial infidelity
  learning for model interpretation,'' in \emph{Proceedings of the 26th ACM
  SIGKDD International Conference on Knowledge Discovery \& Data Mining (KDD),
  Virtual Event}, 2020, pp. 286--296.

\bibitem{zafar2021lack}
M.~B. Zafar, M.~Donini, D.~Slack, C.~Archambeau, S.~Das, and K.~Kenthapadi,
  ``On the lack of robust interpretability of neural text classifiers,'' in
  \emph{Proceedings of the Association for Computational Linguistics Findings
  (ACL/IJCNLP), Virtual Event}, 2021, pp. 3730--3740.

\bibitem{GCJ}
\url{https://codingcompetitions.withgoogle.com/codejam}, 2022.

\bibitem{DBLP:conf/uss/QuiringMR19}
E.~Quiring, A.~Maier, and K.~Rieck, ``Misleading authorship attribution of
  source code using adversarial learning,'' in \emph{Proceedings of the 28th
  {USENIX} Security Symposium, Santa Clara, CA, USA}.\hskip 1em plus 0.5em
  minus 0.4em\relax {USENIX} Association, 2019, pp. 479--496.

\bibitem{abadi2016tensorflow}
M.~Abadi, P.~Barham, J.~Chen, Z.~Chen, A.~Davis, J.~Dean, M.~Devin,
  S.~Ghemawat, G.~Irving, M.~Isard, M.~Kudlur, J.~Levenberg, R.~Monga,
  S.~Moore, D.~G. Murray, B.~Steiner, P.~A. Tucker, V.~Vasudevan, P.~Warden,
  M.~Wicke, Y.~Yu, and X.~Zheng, ``{TensorFlow}: A system for large-scale
  machine learning,'' in \emph{Proceedings of the 12th USENIX Symposium on
  Operating Systems Design and Implementation (OSDI), Savannah, GA, USA}.\hskip
  1em plus 0.5em minus 0.4em\relax {USENIX} Association, 2016, pp. 265--283.

\bibitem{wang-etal-2021-codet5}
Y.~Wang, W.~Wang, S.~Joty, and S.~C. Hoi, ``Code{T}5: Identifier-aware unified
  pre-trained encoder-decoder models for code understanding and generation,''
  in \emph{Proceedings of the 2021 Conference on Empirical Methods in Natural
  Language Processing (EMNLP), Virtual Event}, 2021, pp. 8696--8708.

\bibitem{feng-etal-2020-codebert}
Z.~Feng, D.~Guo, D.~Tang, N.~Duan, X.~Feng, M.~Gong, L.~Shou, B.~Qin, T.~Liu,
  D.~Jiang, and M.~Zhou, ``Code{BERT}: A pre-trained model for programming and
  natural languages,'' in \emph{Proceedings of Findings of the 2020 Conference
  on Empirical Methods in Natural Language Processing (EMNLP), Virtual Event},
  2020, pp. 1536--1547.

\bibitem{russell2018automated}
R.~Russell, L.~Kim, L.~Hamilton, T.~Lazovich, J.~Harer, O.~Ozdemir,
  P.~Ellingwood, and M.~McConley, ``Automated vulnerability detection in source
  code using deep representation learning,'' in \emph{Proceedings of the 17th
  IEEE International Conference on Machine Learning and Applications (ICMLA),
  Orlando, FL, USA}.\hskip 1em plus 0.5em minus 0.4em\relax IEEE, 2018, pp.
  757--762.

\bibitem{alon2019code2vec}
U.~Alon, M.~Zilberstein, O.~Levy, and E.~Yahav, ``code2vec: Learning
  distributed representations of code,'' \emph{Proc. ACM Program. Lang.},
  vol.~3, no. POPL, pp. 1--29, 2019.

\bibitem{allamanis2017learning}
M.~Allamanis, M.~Brockschmidt, and M.~Khademi, ``Learning to represent programs
  with graphs,'' in \emph{Proceedings of the 6th International Conference on
  Learning Representations (ICLR), Vancouver, BC, Canada}, 2018.

\bibitem{puri2017magix}
N.~Puri, P.~Gupta, P.~Agarwal, S.~Verma, and B.~Krishnamurthy, ``{MAGIX}: Model
  agnostic globally interpretable explanations,'' \emph{arXiv preprint
  arXiv:1706.07160}, 2017.

\bibitem{wang2019deepvid}
J.~Wang, L.~Gou, W.~Zhang, H.~Yang, and H.-W. Shen, ``{DeepVID}: Deep visual
  interpretation and diagnosis for image classifiers via knowledge
  distillation,'' \emph{IEEE Transactions on Visualization and Computer
  Graphics}, vol.~25, no.~6, pp. 2168--2180, 2019.

\bibitem{simonyan2013deep}
K.~Simonyan, A.~Vedaldi, and A.~Zisserman, ``Deep inside convolutional
  networks: Visualising image classification models and saliency maps,'' in
  \emph{Proceedings of the 2nd International Conference on Learning
  Representations (ICLR), Banff, AB, Canada}, 2014.

\bibitem{lundberg2017unified}
S.~M. Lundberg and S.-I. Lee, ``A unified approach to interpreting model
  predictions,'' in \emph{Proceedings of Annual Conference on Neural
  Information Processing Systems (NeurIPS), Long Beach, CA, USA}, 2017, pp.
  4765--4774.

\bibitem{schwab2019cxplain}
P.~Schwab and W.~Karlen, ``{CXPlain}: Causal explanations for model
  interpretation under uncertainty,'' in \emph{Proceedings of Annual Conference
  on Neural Information Processing Systems (NeurIPS), Vancouver, BC, Canada},
  2019, pp. 10\,220--10\,230.

\bibitem{smilkov2017smoothgrad}
D.~Smilkov, N.~Thorat, B.~Kim, F.~Vi{\'e}gas, and M.~Wattenberg,
  ``{SmoothGrad}: Removing noise by adding noise,'' \emph{arXiv preprint
  arXiv:1706.03825}, 2017.

\bibitem{rieger2020simple}
L.~Rieger and L.~K. Hansen, ``A simple defense against adversarial attacks on
  heatmap explanations,'' in \emph{Proceedings of the 2020 Workshop on Human
  Interpretability in Machine Learning (WHI), Virtual Event}, 2020.

\bibitem{wang2020smoothed}
Z.~Wang, H.~Wang, S.~Ramkumar, P.~Mardziel, M.~Fredrikson, and A.~Datta,
  ``Smoothed geometry for robust attribution,'' in \emph{Proceedings of Annual
  Conference on Neural Information Processing Systems (NeurIPS), Virtual
  Event}, 2020, pp. 13\,623--13\,634.

\bibitem{zhao2021baylime}
X.~Zhao, W.~Huang, X.~Huang, V.~Robu, and D.~Flynn, ``{BayLIME}: Bayesian local
  interpretable model-agnostic explanations,'' in \emph{Proceedings of the 37th
  Conference on Uncertainty in Artificial Intelligence (UAI), Virtual Event},
  2021, pp. 887--896.

\bibitem{zhou2021s}
Z.~Zhou, G.~Hooker, and F.~Wang, ``{S-LIME}: Stabilized-lime for model
  explanation,'' in \emph{Proceedings of the 27th ACM SIGKDD Conference on
  Knowledge Discovery \& Data Mining (KDD), Virtual Event}, 2021, pp.
  2429--2438.

\end{thebibliography}

\end{document}